\documentclass[10pt, preprint2]{aastex}
\begin{document}


\title{MEASURING THE 3D SHAPE OF X-RAY CLUSTERS}


\author{Johan Samsing\altaffilmark{1},  Andreas Skielboe\altaffilmark{1}, Steen H. Hansen\altaffilmark{1}}
\affil{ \altaffilmark{1} Dark Cosmology Centre, Niels Bohr Institute, University of Copenhagen, Juliane Maries Vej 30, 2100 Copenhagen, Denmark}



\begin{abstract}
Observations and numerical simulations of galaxy clusters strongly indicate that the hot intracluster x-ray emitting gas is not spherically symmetric. 
In many earlier studies spherical symmetry has been assumed partly because of limited data quality, however new deep observations and instrumental designs will make it possible to go 
beyond that assumption. Measuring the temperature and density profiles are of interest when observing the x-ray gas, however the spatial shape of the gas itself also carries 
very useful information. For example, it is believed that the x-ray gas shape in the inner parts of galaxy clusters is greatly affected by feedback mechanisms, cooling 
and rotation, and measuring this shape can therefore indirectly provide information on these mechanisms. 
In this paper we present a novel method to measure the three-dimensional shape of the intracluster x-ray emitting gas. We can measure the shape from the x-ray observations only,
i.e. the method does not require combination with independent measurements of e.g. the cluster mass or density profile. This is possible when one uses the full spectral information
contained in the observed spectra. We demonstrate the method by measuring 
radial dependent shapes along the line of sight for \verb+CHANDRA+ mock data. 
We find that at least $10^{6}$ photons are required to get a $5-\sigma$ detection of shape for an x-ray gas having realistic features such as a cool core and a double powerlaw 
for the density profile. We illustrate how Bayes' theorem is used to find the best fitting model of the x-ray gas, an analysis that is very important in
 a real observational scenario where the true spatial shape is unknown. Not including a shape in the fit may propagate to a mass bias if the x-ray is used to estimate the
 total cluster mass. We discuss this mass bias for a class of spacial shapes.  
\vspace{0.5in}

\end{abstract}

\section{INTRODUCTION}

Galaxy clusters are the largest bound objects in the universe and they provide unique and independent information on the cosmological
evolution. The standard LCDM parameters and a possible redshift varying dark energy component has accurately been measured and constrained from cluster observations in a variety of 
ways (\cite{2009astro2010S.305V}, \cite{2011ARA&A..49..409A}, \cite{2010MNRAS.406.1759M}, \cite{2008MNRAS.383..879A}, \cite{2009ApJ...692.1060V}). 
They reveal the distant universe behind them through gravitational
 magnification (\cite{2004ApJ...607..697K}, \cite{2011arXiv1109.4740A}, \cite{2011arXiv1104.2035B}), and they are even sensitive to 
the initial perturbations of our universe (\cite{2009MNRAS.397.1125F}, \cite{2011arXiv1107.5617C},  \cite{2010MNRAS.407.2339S}).
Clusters not only serve as excellent laboratories for constraining the standard cosmology, but because of their relative high mass and cosmological 
size they also provide a unique possibility to test general relativity itself in several independent ways, e.g. from measurements of 
cosmic growth (\cite{2010MNRAS.406.1796R}) to gravitational redshift (\cite{2011arXiv1109.6571W}) 
and gravitational waves (\cite{2009PhRvD..80h3514Y}). Other probes have also been suggested such as lensing, cluster abundance and
 the integrated Sachs-Wolfe effect (\cite{2008PhRvD..78f3503J}). 
Despite their importance in modern cosmology, basic
properties such as spatial shape is still not well measured for individual
clusters. One reason is simply that the main part of a cluster is
composed of dark matter which can only be measured indirectly by its
gravitational interaction. The indirect measurements of the dark matter
and its radial distribution are usually done using either lensing
(\cite{2011arXiv1106.3328P}, \cite{2007ApJ...663...10S}), by studying the dynamics of the 
intracluster galaxies (\cite{2010MNRAS.408.2442W}, \cite{2003MNRAS.343..401L}, \cite{2009ApJ...701.1336L})
or by the hot baryonic x-ray emitting gas located in the inner regions
of all clusters (for a review of x-ray physics and applications see e.g. \cite{1988xrec.book.....S}). Especially 
observations of the intracluster x-ray gas in terms of spacial shape, density and temperature profiles, play
a key role for estimating local properties of the cluster. Many earlier
studies assume a spherical shape of the gas 
(\cite{2005A&A...435....1P}, \cite{2011ApJ...736...52H},  \cite{2004A&A...413..415K},  \cite{2005A&A...433..101P},  \cite{2007A&A...476L..37H},  \cite{2010MNRAS.406.1796R}), 
however there are several
strong motivations why a precise estimation of the shape is interesting.
One is a precise estimation of the cluster mass profile. This profile
can directly be measured if the radial shape, temperature and density
profiles of the gas are known and the gas is in hydrostatic equilibrium. Only recently
 it was shown that allowing the gas to have
a triaxial shape is necessary for the estimated mass profile from
x-ray to agree with the mass estimated from lensing 
(\cite{2010ApJ...713..491M}, \cite{2011MNRAS.416.3187S}, \cite{2011MNRAS.416.2567M}), a result
 in good agreement with numerical simulations (\cite{2007MNRAS.377...50H}).
 This overall triaxiality is mostly due to the underlying shape of the dark matter potential. However, in the central cluster regions
 it is believed that a possible non-spherical x-ray shape is more affected by
microphysical processes such as radiative cooling, turbulence and
different feedback mechanisms (\cite{2011ApJ...734...93L}) than the dark matter potential shape is. These mechanisms change the gas
shape into having relatively high ellipticity towards the center compared
to the underlying dark matter potential shape. It is therefore
possible to infer properties of these mechanisms if the shape of the
gas, temperature and density profiles are known to high precision. 

In this paper we suggest and develop a method from which a possible radial
dependent shape of an x-ray gas can be extracted from the x-ray observations
only. We explicitly demonstrate the possibilities for measuring the shape
by fitting to \verb+CHANDRA+ mock data and we estimate the mass bias if a shape is
not treated correctly in the fitting. The method we use is a parametrized
approach, i.e. we assume that the shape and profiles can be described
by a set of well defined functional forms. We also discuss the complications
of choosing the best set of functions, i.e. a model, to describe the
data. 

The paper is organized in the following way; The method for measuring shape is explained in section 
\ref{intro2method}. We apply the method in section \ref{results_section} on \verb+CHANDRA+ mock data.
We discuss how to quantify the goodness of fit in section \ref{sub:Quantifying-the-goodness}.
Mass bias from not including the shape in the fitting is discussed in section \ref{massbias_section}.

\section{EXTRACTING 3D X-RAY INFORMATION FROM 2D OBSERVATIONS\label{intro2extracting3dinfo}}

\begin{figure}[h!]
\center
\includegraphics[scale=0.42]{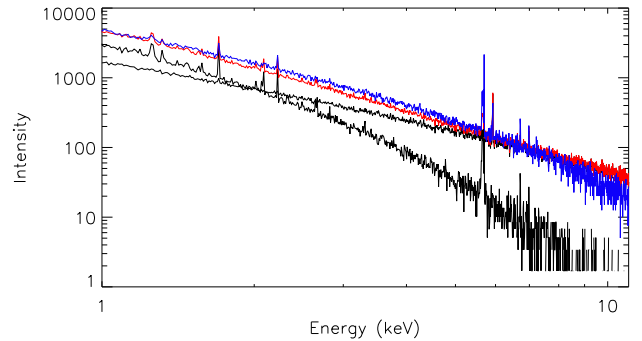}
\caption{\small{Simplified description of how information propagates from three dimensions
to a two dimensional observation. The figure shows two free-free emission
X-ray spectra (the two lowest spectra at 3 keV in black), the sum of these two spectra (second spectrum from the top at 3 keV in red),
and a best fit free-free spectrum to the red spectrum (upper spectrum at 3 keV in blue).
This could correspond to a part of a gas with two temperature components
(one spectrum for each component) projected along the line of sight
(observed data). By comparing the blue and the red spectra one notices
that the red spectrum cannot simply be fitted accurately with a free-free
emission X-ray spectrum. This is true in the general case; a sum of
free-free spectra cannot in general be fitted by another free-free
spectrum. If the observed red spectrum is not correctly fitted one
could incorrectly conclude the presence of either a \emph{non thermal}
\emph{hard X-ray excess} or \emph{non thermal} \emph{soft X-ray excess}
component, depending on the shape of the red spectrum. In our example
one would report a \emph{hard excess} component since the red spectrum
is above the blue in the tail (see e.g. \cite{2005ApJ...624L..69F} for a hard excess discussion for Abell 2256).
 \label{fig:specinfo_example}}}
\end{figure}

An intracluster x-ray emitting gas has a three dimensional extension,
spherical or not, but an observer will only see the two dimensional
projected image on the sky. Therefore, a given observed spectrum is
a sum of all emission spectra along the line of sight through the
gas (for a discussion see figure \ref{fig:specinfo_example}).

Each spectrum has a spectral shape determined by the local temperature
and a scaling proportional to the local density squared (\cite{1988xrec.book.....S}).
Mathematically, no unique mapping can construct the true three dimensional
shape, density and temperature profiles using only the observed two
dimensional image. However, if one makes prior assumptions it can
be done. For instance by assuming that the gas is spherical the density
and temperature profiles can be found. From this assumption several previous 
groups have measured the temperature and density profiles of the
x-ray gas using either projection or de-projection techniques (see e.g. the \verb+XSPEC+ packages 'deproject' and 'projct')

The method we present in this paper for extracting three-dimensional information
relies on the assumption that the x-ray gas shape,
density and temperature profiles can be described by parametrizations.
This means that shape and profiles are believed to be well described
by a set of functions. In contrast to several previous studies we use the whole spectral information from the integrated observed picture of the x-ray
gas. It means that we take into account that the actual observed spectra is a sum of spectra along the line of sight, and can not
simply be fitted by a single free-free spectrum. See figure \ref{fig:specinfo_example} for a discussion.  
We allow a radial dependent shape in contrast to previous studies.
The tradeoff for including this extra freedom is that we limit our
analysis to structures that are seen spherical on the sky. This is for purely practical reasons: in theory 
the fitting method we describe is not limited by this assumption, but with present day available data it is simply not possible
 to resolve a radial dependent shape if the 3D-shape and orientation of the gas is completely free to vary. In other words, the symmetry in the sky 
makes it possible to extract higher order corrections to the usual assumption about either sphericity or triaxiality with constant axis ratios. 
The method and procedure will be described in the following sections, and technical
details are found in the appendix together with illustrations of generated spectra and an x-ray structure.

\subsection{Fitting shape and profiles using the parameterization approach\label{intro2method}}

The procedure needed in order to measure spatial shape, temperature
and density profiles of an observed x-ray gas using the parametrization
approach is as following: First we choose a model, i.e.
a set of parametrizations, that are believed to generally describe
the form of density, temperature and spatial shape (along the line
of sight) for the observed structure. The chosen parameterizations
must be sufficiently general to accurately describe observations of
real and simulated structures. We then calculate
the agreement between an artificial generated dataset 
(see appendix section \ref{sub:Creating-artificial-observations} for how we generate artificial datasets and mock data)
created from the chosen model given a specific combination of parameter values and the observed dataset.
In our case we quantify the agreement by a simple
$\chi^{2}$ statistic which simply can be related to a probability by $exp(-\chi^{2}/2)$ when the noise is gaussian.
This routine of comparing artificial generated datasets with the observed dataset is then repeated 
for a wide range of parameter value combinations until a good estimate
of the underlying probability distribution function (PDF) for our model has been
made. For this we use standard Monte Carlo techniques as described in
section \ref{sub:Monte-Carlo-Technics} in the appendix. From the
parameter combination having the maximum PDF value, the best estimate
for profiles and shape, given our prior input parameterizations, can
then be made. The overall procedure can then be repeated for different models, until
the best model is found. We will discuss this in more detail in section \ref{sub:Quantifying-the-goodness}.

\section{RESULTS FROM FITTING SHAPE AND PROFILES OF SELECTED X-RAY MODELS\label{results_section}}

In the following we show the possibilities of measuring radial profiles
of non-spherical x-ray structures with varying radial dependent shape
along the line of sight. 
As briefly discussed in the end of section \ref{intro2extracting3dinfo}, we only consider structures that are spherical on the sky.
We consider two simulated structures in our
analysis; First a simple toy model to clearly illustrate the method, and second
a more realistic model with features such as a cool core and a double
powerlaw for the density profile. The shape parameterizations are described
later. In this part of the analysis we fit for profiles and shape
using the same set of parameterizations that are used to generate the
data. In this way we get the cleanest picture of how a shape signal
propagates to observables. 

We present results in terms of a virial radius $r_v$. The shape, temperature and 
density profiles we use, are consistent with a virial radius similiar to $r_{500}$ (\cite{2006ApJ...640..691V}).

\subsection{A simple toy model \label{sub:A-simple-toy}}

\begin{figure*}
\center
\includegraphics[scale=0.85]{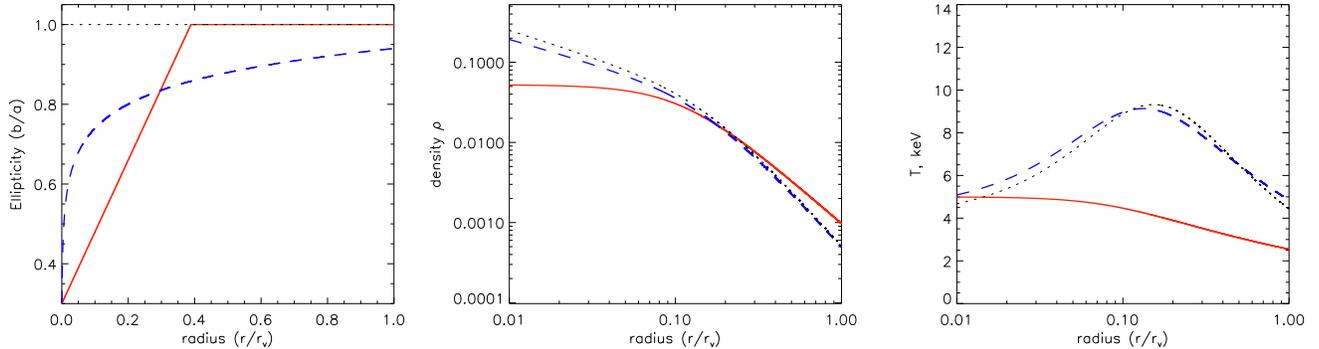}
\caption{\small{Shape and profiles for dataset shM1 (red, solid lines) and shM2 (blue, dashed lines).
 Left: Ellipticity of the x-ray gas along the line
of sight. Central: x-ray gas density profile. Right:
x-ray gas temperature profile. Black, dotted lines: Best estimate
from a MCMC fitting to shM2 using the true parametrizations for temperature
and density, but assuming spherical symmetry.\label{fig:Shape-and-profiles}}}
\end{figure*}

We consider a dataset denoted by 'shM1' where the density and temperature
profiles are modeled by simple broken powerlaws

\begin{equation}
\rho(r)=n_{0}(1+(r/r_{c})^{2})^{-3{\beta}/2}\label{eq:10}\end{equation}

\begin{equation}
T(r)=T_{0}(1+(r/r_{t})^{2})^{-b}\label{eq:20}\end{equation}
known as beta-models. Parameter $n_{0}$ acts as a normalization
factor and is regulated such that the artificial dataset has a fixed number
of total (photon) counts. The shape parametrization we consider is
a simple linear function for ellipticity

\begin{equation}
\epsilon_{1}(r)=s_{2}\cdot r+s_{1}\label{eq:30}\end{equation}
where $\epsilon\equiv b/a$ is defined as the ratio between the radius
perpendicular to the observer ($b$) and the radius along the line
of sight ($a$) of the observer. The parameter values for shM1 are
listed in table \ref{tab:info spM1 shM1}, and figure \ref{fig:Shape-and-profiles} shows the corresponding
shape and profiles. The chosen parameters for the density
and temperature profiles are in fair agreement with typical observed
values. The priors on the shape parametrization we use in this example are: a) $0.2<b/a\leq1$ and b) $\epsilon_{1}(r/r_{v}=1)>0.5$.
In general, a structure could naturally have an axis ratio $b/a\geq1$ and still be spherical on the sky, and therefore in a scenario
where no prior shape information is available, shapes with $b/a\geq1$ must be included in the fit as well.

\begin{deluxetable}{lrrrrcrrrrrr}
\tablewidth{0pt}
\tablecaption{Simulated profiles\label{tab:info spM1 shM1}}
\tablehead{
\colhead{Model}           & \colhead{equation}      &
\colhead{$n_{0}$}          & \colhead{$r_{c}$}  &
\colhead{$\beta$}          & \colhead{${\alpha}/2$}    &
\colhead{$T_{0}$}  & \colhead{a}  &
\colhead{b} & \colhead{$r_{t}$} &
\colhead{$s_{1}$} & \colhead{$s_{2}$}}
\startdata
shM1 & 1,2,3 & \nodata & 0.11 & 0.6 & 0 & 5.0 & 0 & 0.14 & 0.09 & 0.3  & 0.83 \\
shM2 & 4,6,7 & \nodata & 0.15 & 0.76 & 1.2 & 4.3 & 2.45 & 0.7 & 0.13 & 0.94 & 0.2 \\
\enddata
\end{deluxetable}

The left plot of figure \ref{fig:PDFs-rho,t} shows the maximized
PDFs for the fitted density, temperature and shape parameters for
a total of $3\cdot10^{4}$ photon counts ($\approx$ 10 ks \verb+CHANDRA+ observation of A1689). The width of the projected
PDFs, i.e. a measure of the fitting error for each parameter, is simply
related to the number counts by $\sim1/\sqrt{N}$ where $N$ is the
number of photons. The right plot of figure \ref{fig:PDFs-rho,t}
shows the corresponding correlation matrix defined in the usual way
as 
$CORR(X,Y)=\frac{\left\langle (X-\mu_{X})(Y-\mu_{Y})\right\rangle }{\sigma_{X}\sigma_{Y}}$
where $X,Y$ are random variables with expectation values $\mu_{X},\mu_{Y}$
and standard deviations $\sigma_{X},\sigma_{Y}$. In our case, to
find e.g. the correlation coefficient $CORR(p_{i},p_{j})$ $X$ must
be replaced with the vector of MCMC sampled $p_{i}$ values and $Y$
of sampled $p_{j}$ values. The correlation matrix is symmetric by
construction and the shading goes from $0$ (black) to $1$ (white). The correlation matrix can be divided up in several
regions. On the plot is highlighted a region bounded by a dotted line
and a solid line. The dotted region is the part which shows the correlation
between temperature and density and the solid is the region that shows
the correlation between temperature and shape. In the lower left corner
of the correlation matrix is the region showing the correlation between
shape and density. In general we see that the temperature is weakly
correlated with the rest of the parameters, especially when compared
to the correlation between shape and density. The physical reason is simply that their individual contributions
to a spectrum by nature are completely different; temperature
affects the spectral form, but density and shape affect only the normalization. This is clearly
seen in the analytic form for the bremsstrahlung spectrum $I(T,\rho,\nu)\propto\rho^{2}T^{-1/2}exp(-\nu/T)$
(\cite{1988xrec.book.....S}).

Among our chosen priors, the prior on $s_{1}$ ($s_{1}>0.2$) is the
one that affects the shape of the PDFs the most. Besides a trivial
truncation on the $s_{1}$ parameter axis it is also responsible for
especially the truncation (or skewness) of the $n_{0}$ distribution.
The reason is the relative strong correlation between these two parameters.
This correlation is clearly seen on the correlation matrix and can
be understood in the following way: The degree of constant ellipticity
captured by $s_{1}$ effectively acts as mass scaling term when the
structure is projected along the line of sight. This is simply because
an ellipticity {}``stretches'' the structure and therefore {}``allows''
more mass along the line of sight. This is exactly how $n_{0}$ affects
the projected dataset too. So if we increase the overall scaling (increasing
$n_{0}$) we can compensate by decreasing the ellipticity (increasing
$s_{1}$), that means the lower truncation of $s_{1}$ also shows
up as a lower truncation on $n_{0}$. In fact, a constant ellipticity along the line of sight
$\epsilon$ is completely degenerate with the overall density scaling
$\rho_{0}$ by $\rho_{0}^{2}\epsilon$. This is an intrinsic degeneracy
and can only be broken by including other observations, e.g. SZ observations
which effectively traces $\rho_{0}{\epsilon}T$ (see e.g. \cite{2011arXiv1101.2043P}, \cite{2005ApJ...625..108D}, \cite{2011A&A...532A..14C}, \cite{2011arXiv1109.2732S}).

The overall conclusion from the fitting results is that the parameter
values specifying the true shape as well as temperature and density are exactly
reconstructed. This is an ideal case, but it is clearly showing that
temperature, density and shape in principle can be separated. 

From the correlations we can conclude that the temperature profile
is well and almost independently fitted. In perspective of optimizing
the fit for shape, this also implies that independent measurements
of the density will directly result in a better fit for the shape.

\begin{figure*}
\center
\includegraphics[scale=0.49]{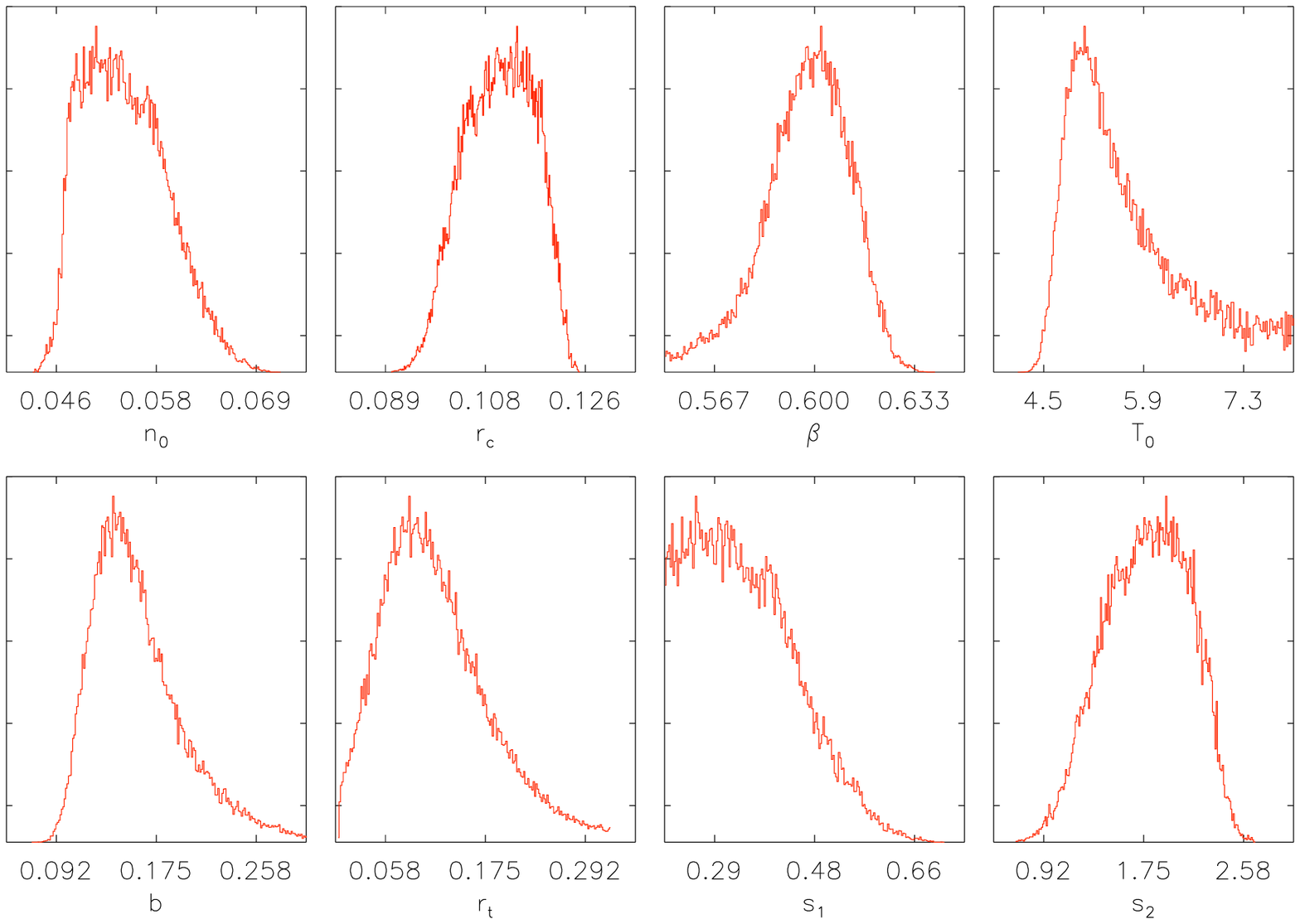}\includegraphics[scale=0.80]{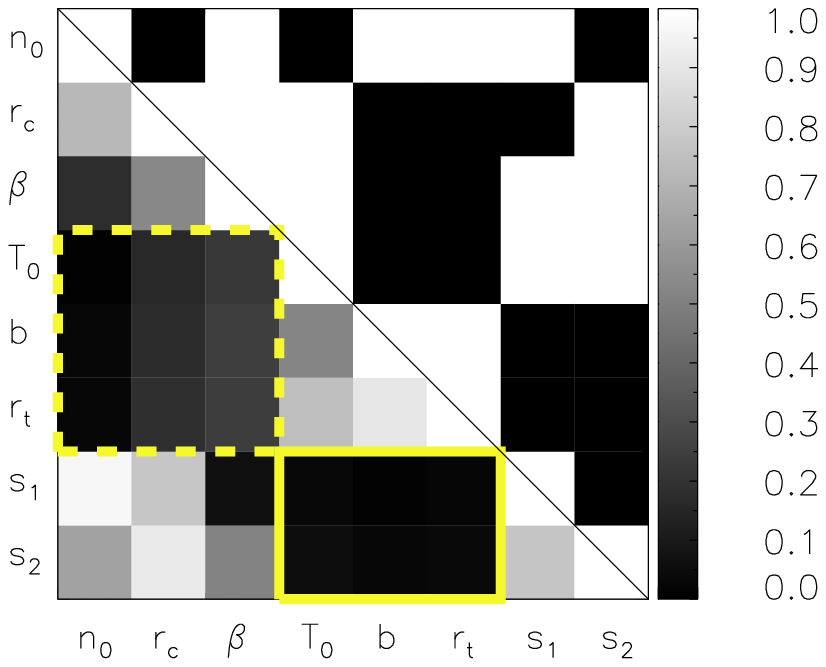}

\caption{\small{Fitting results for dataset 'shM1'. Left: Maximized 
PDFs along each of the 8 parameters in the model. Right:
Correlation matrix for the 8 parameters in the model. The lower left part of the matrix shows the absolute value of the correlation coefficient 
where the upper right corner shows the sign of the coefficient in black (negative) and white (positive).
The results are based on a dataset scaled to have a total of $3\cdot10^{4}$ photon
counts. \label{fig:PDFs-rho,t}}}

\end{figure*}

\subsection{A more realistic model\label{sub:A-more-realistic}}

We now perform an analysis on a dataset, denoted by 'shM2', describing
a structure with cool core and a double powerlaw for the density profile.
Including these features are motivated by real observations (\cite{2006ApJ...640..691V}).
The temperature and density profiles are now parameterized as 

\begin{equation}
\rho(r)=n_{0}\frac{(r/r_{c})^{-{\alpha}/2}}{(1+(r/r_{c})^{2})^{3{\beta}/2-{\alpha}/4}}\label{eq:50}\end{equation}
\begin{equation}
T(r)=T_{0}\frac{1+a(r/r_{t})}{(1+(r/r_{t})^{2})^{b}},\label{eq:60}\end{equation}
and the shape is parameterized by

\begin{equation}
\epsilon_{2}(r)=s_{2}\cdot log_{10}(r)+s_{1}\label{eq:40}\end{equation}
This shape parametrization approximately describes the gas shape seen
in the inner parts $(r\leq r_{500})$ of clusters in numerical simulations
(\cite{2011ApJ...734...93L}). We use the same shape priors
as used in the previous toy model example. A list of temperature and density parameterizations
are found in (\cite{2006ApJ...640..691V}).

The true parameter values for 'shM2' are listed in table \ref{tab:info spM1 shM1}
and the corresponding shape and profiles are plotted in figure \ref{fig:Shape-and-profiles}.
Figure \ref{fig:PDFs-rho,t_compl1} shows the PDF and
the correlation matrix for the 10 parameter model fitting. 

An inner density slope captured by ${\alpha}$ is now one of the new
parameters compared to the toy model. Since both the shape and 
this inner slope have a logarithmic dependence,
there is a strong correlation between ${\alpha}$ and $s_{2}$. This
is clearly seen in the correlation matrix and the PDF plot where the lower cut on $s_{2}$
directly relates to the skewness in the ${\alpha}$ distribution. This
freedom in the inner slope is the main reason for the fitting to require
many more photons than the toy model. This is discussed in more detail
in section \ref{sub:Quantifying-the-goodness} below.

The overall conclusion is that the true parameter values are reconstructed,
but to keep down the statistical errors a relative high number of photons are
required. This is mostly due to the similar parameterizations for shape and density. 
In agreement with intuition, we see that it is much harder to extract a logarithmic shape
when the density is varying logarithmically too, compared to e.g. a linear dependent shape.
From the correlation matrix we see that the temperature fitting is nearly
unaffected as we also concluded in the previous toy model example.

\begin{figure*}
\center
\includegraphics[scale=0.49]{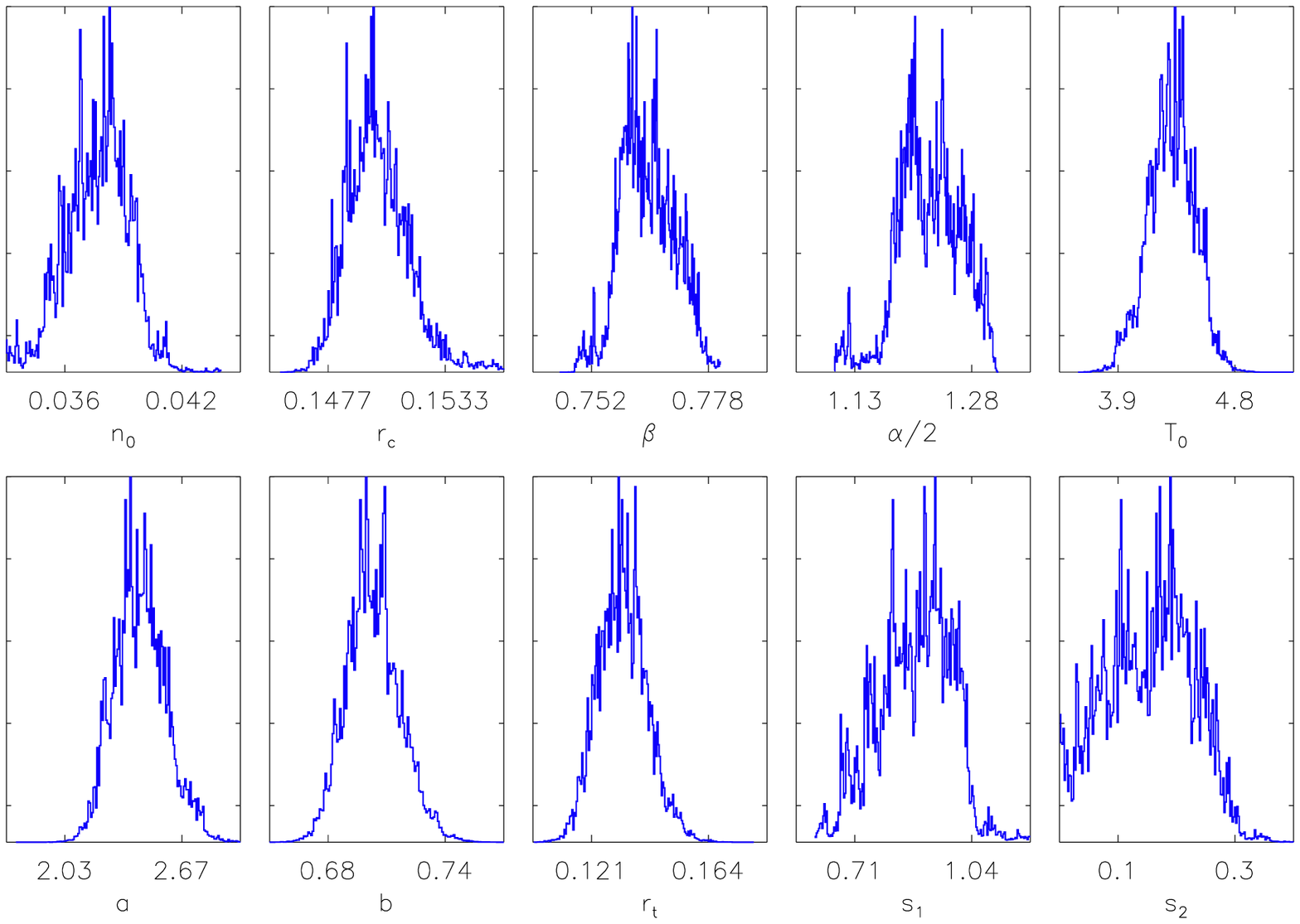}\includegraphics[scale=0.80]{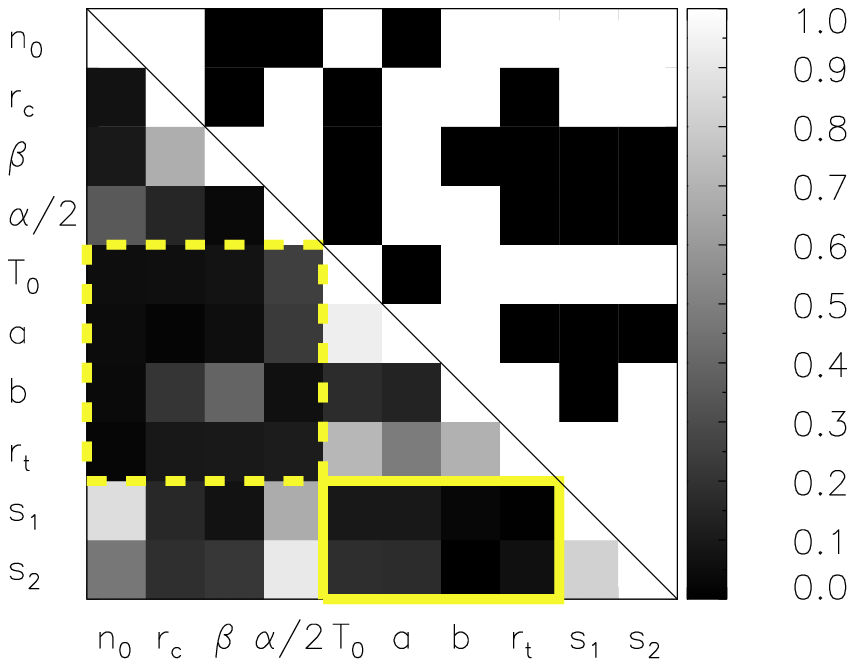}

\caption{\small{Fitting results for dataset 'shM2'. Left: Maximized 
PDFs along each of the 10 parameters in the model. Right:
Correlation matrix for the 10 parameters in the model. 
The lower left part of the matrix shows the absolute value of the correlation coefficient 
where the upper right corner shows the sign of the coefficient in black (negative) and white (positive).
The results
are based on a dataset scaled to have a total of $1\cdot10^{6}$ photon
counts. \label{fig:PDFs-rho,t_compl1}}}

\end{figure*}

\subsection{Quantifying the goodness of fit\label{sub:Quantifying-the-goodness}}
In this section we will discuss how to quantify the goodness of fit for the parameters within a given model, as well as the goodness of fit for the model itself relative to other competitive models.
\subsubsection{Individual parameters within a model\label{sub:Individual parameters fit}}
\begin{figure*}
\center
\includegraphics[scale=0.50]{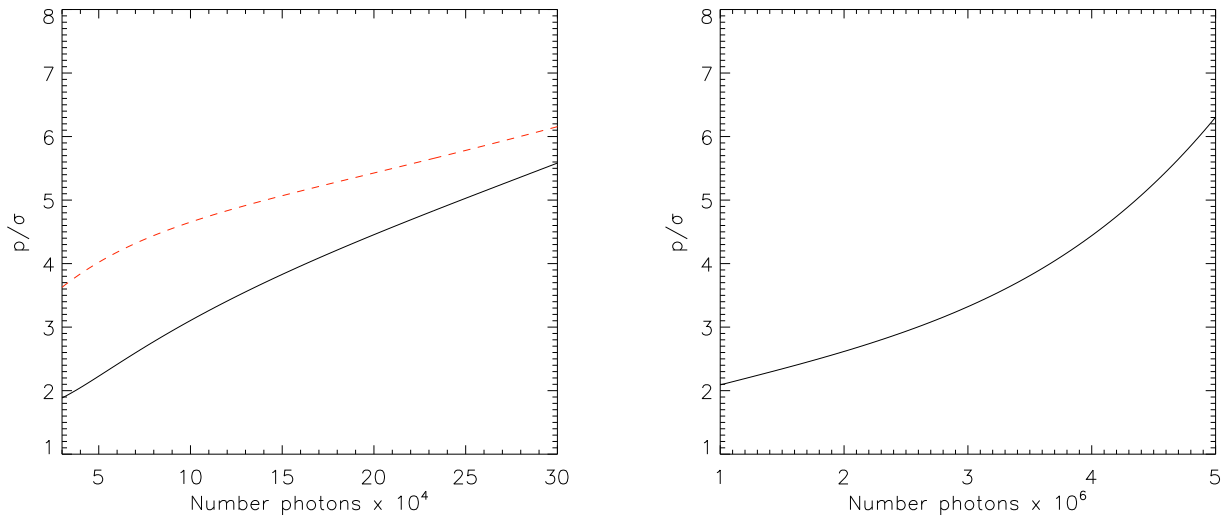}

\caption{\small{The plot shows the ratio $n=\hat{p}_{i}/\sigma_{i}$ for the 
most difficult parameters to estimate when fitting for structure 'shM1' (left figure) and 'shM2' (right figure) in black lines. The worst estimated parameter in terms of $n$
when fitting to 'shM1' is $r_{t}$ and $s_{2}$ when fitting to 'shM2'. The red dashed line in the left plot shows the ratio for parameter $s_{1}$ which is the worst determined of the two shape parameters when fitting
to 'shM1'. It can be read of the figure that $\approx4.2\cdot10^{6}$ number total photon counts
are required for a minimum $5-\sigma$ detection on all parameters when fitting to 'shM2'. The radial and spectral binning is kept constant in the plot.
\label{fig:relative_sigma_plot}}}

\end{figure*}

The best fit parameter values for a given model are located at the
likelihood maximum, or the minimum $\chi^{2}$ if the measurement
noise is gaussian. To quantify the goodness of the fit is not unique
in the same way. To quantify this one must often combine statistical
estimators with prior knowledge. An often used estimator is the reduced
chi square, $\chi_{red}^{2}=\chi^{2}/K$, where $K$ is the number
of degrees of freedom. However, this estimator has two major problems.
First, $\chi^{2}$ itself have a significant noise due to random noise
of the data, and second, the number of degrees of freedom is not in general
well defined (\cite{2010arXiv1012.3754A}). Another, maybe more intuitive, estimator is
the the ratio $\hat{p}_{i}/\sigma_{i}$ where $\hat{p}_{i}$ is the
best estimate for parameter $p_{i}$ and $\sigma_{i}$ the associated
standard deviation. If we denote this ratio by $n$ we can quantify
the goodness of fit by reporting $n$ for each parameter or the minimum
$n$ for the whole model. For the fitting examples we presented above,
it is then of interest to know the number of photons required for
e.g. a minimum $n=5$ (or $5-\sigma$) detection for all parameters.
We will investigate this in the following. 
Figure \ref{fig:relative_sigma_plot} shows the ratio $\hat{p}_{i}/\sigma_{i}$ as a function of total photon counts for the parameter that 
have the largest ratio, i.e. the parameter which is most difficult to estimate, for the two structures 'shM1' (left plot) and 'shM2' (right plot).
In the 'shM1' example the most difficult parameter to estimate in terms of $n$ is $r_{t}$, to reach a minimum $5-\sigma$ detection
of this (and thereby for each parameter of the whole model) we find from the figure that 
more than $\approx2.5\cdot10^{5}$ photons are required. If we instead only 
require that the shape parameters must be estimated with a minimum
$5-\sigma$ each we find a limit of $\approx1.4\cdot10^{5}$ photons,
or roughly a factor of $2$ less compared to an overall $5-\sigma$
detection. Following the same procedure for the more realistic example 'shM2'
we find that a minimum of $\approx4.2\cdot10^{6}$  photons
are required for a minimum $5-\sigma$ detection on all parameters.
The same number of photons are required for the shape fitting because
$s_{2}$ is the most difficult parameter to estimate in terms of $n$.

\subsubsection{Model comparison\label{sub:Model comparison fit}}
Assuming that the quantities we try to measure for a gas can be parameterized,
we still have the problem that we have no idea of how the {}``true''
or best parametrization for the gas looks like in a real observation.
This means e.g. that a set of shape parameters defined in a specific
gas parametrization do not have to describe a real shape at all. The
parameters could in principle just capture higher order corrections
to the density profile, because of the general tight correlation between
shape and density. In this case, the real problem is to realize that
your model does not return information about the system in the way you
believe. The question is therefore how to quantify how a specific model
performs relative to one or several other competitive models. A useful
measure of this can be found using Bayes' theorem. From this theorem
it is possible to calculate the relative probability, also known as
the \emph{posterior odds,} of two competing models (\cite{2011MNRAS.413.2895J}, \cite{2008ConPh..49...71T}). In the
case where we assume flat parameter and model priors, the posterior odds
ratio reduces to the simple ratio

\begin{equation}
\mathcal{F}(H_{1},H_{0})=\int\mathcal{L}(D\mid H_{1},\beta)d\beta/\int\mathcal{L}(D\mid H_{0},\alpha)d\alpha\label{eq:evidence ratio}\end{equation}
where $\mathcal{L}(D\mid H,\beta)$ is the likelihood for getting
the data $D$ given the model $H$ which depends on the parameterset
$\beta$. This ratio $\mathcal{F}$ is often denoted the \emph{evidence
ratio} between model $H_{1}$ and $H_{0}$. Model $H_{0}$ is often
a 'null' or default model where $H_{1}$ is a competing and often
more complicated model. In our case, $H_{0}$ could be a model assuming
spherical symmetry and $H_{1}$ a model allowing the shape to vary.
The evidence threshold, or critical threshold, between rejecting or
accepting a competitive model is often taken to be Jeffreys threshold
1:148 (\cite{2011MNRAS.413.2895J}). Let us now go through a few examples.

First suppose 
we want to compare two models,
$M_{1}$ and $M_{2}$, given the data set shM2. Both models are using
the correct parameterizations for temperature and density, but not
the same parametrization for shape; model $M_{1}$ includes the true
parametrization of shape in the fitting, but model $M_{2}$ assumes
spherical symmetry. We can now use Bayes' theorem to show if e.g. a
$5\cdot10^{5}$ photon exposure carries enough information to distinguish
between $M_{1}$ and $M_{2}$. Performing the two integrals in equation \ref{eq:evidence ratio}
for a $5\cdot10^{5}$ photon exposure we find $\mathcal{F}\thicksim900$,
i.e. we can correctly conclude that $M_{1}$ is strongly favored over $M_{2}$. The slightly biased
estimations for the density and temperature when shM2 is fitted assuming
$M_{2}$ is seen in figure \ref{fig:Shape-and-profiles}.

Another scenario could be that we fit the shape with a parametrization
that is different from the true one. In that case, suppose we fit
dataset shM1 with two models $M_{1}$ and $M_{2}$. Both of them are
using the true temperature and density parameterizations, but model
$M_{1}$ is using equation \ref{eq:30} for the shape parameterization
in contrast to model $M_{2}$ that is using equation \ref{eq:40}.
For a $3\cdot10^{4}$ photon exposure we find $\mathcal{F}\thicksim6700$,
concluding correctly that $M_{1}$ is strongly favored over $M_{2}$. 

\begin{figure}[h!]
\centering
\includegraphics[scale=0.9, viewport= 10 150 270 340 ]{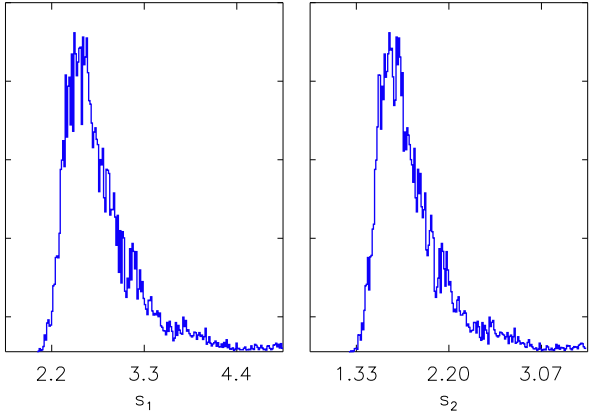}
\caption{\small{Example of a false detection of shape. Maximized PDFs for the two shape parameters $s_{1}$ and $s_{2}$ when fitting to a spherical version of  'shM2', but 
using a beta-model for the density instead of the true equation \ref{eq:50}. The peaks in the shape parameters are real, but are not representing shape. They reflect 
that the parameterization for density used in the fitting is not having enough freedom to describe the variation of density in the inner parts. As described in the text,
Bayes' theorem can be used to quantify if this is a true signal of shape or not. The plot is for a $6\cdot10^{4}$ photon exposure.
\label{fig:falsedetec}}}
\end{figure}

The last example is a case where the true structure has temperature and density profiles as 'shM2',
but have a spherical shape. We now make a fit including shape, but we use equation \ref{eq:10}, i.e. a simple beta-model,  to describe the density instead
of the true equation \ref{eq:50} that has one extra degree of freedom. The interesting thing is now that the best fit using the beta-model will show clear 
detection of shape away from spherical. This is seen on figure \ref{fig:falsedetec}.
The under fitted density profile is simply compensated by allowing a non-spherical shape in the inner parts. 
This is a false detection. In a real case where the true shape of the gas is not known, this can be very hard to realize. Comparing this fit using 
Bayes' theorem with a fit using the more general density profile in equation \ref{eq:50} we find $\mathcal{F}\thicksim300$ for a $6\cdot10^{4}$ photon exposure. 
Which correctly means a spherical model is favored. 

It is possible to write up a simple scaling relation between number photons and the evidence ratio given that the PDF
approximately can be described by a multidimensional gaussian near its peak; Assume from a $N_{2}$ photon exposure we have 
calculated the evidence ratio $\mathcal{F}_{N_{2}}$ between two models $M_{A}$ and $M_{B}$, from that we can 
simply calculate the ratio $\mathcal{F}_{N_{1}}$ for a $N_{1}$ photon exposure by
 $\mathcal{F}_{N_{1}}{\approx}\mathcal{F}_{N_{2}}(P^{*}_{M_{A}}/P^{*}_{M_{B}})^{(N_{1}/N_{2}-1)}$ where 
$P^{*}$ is the value of the PDF at its maximum for the $N_{2}$ photon exposure. Here we have used the analytical solution 
to equation \ref{eq:evidence ratio} (see e.g. \cite{2011MNRAS.413.2895J} eq. 8). 
This scaling relation can be useful for forecasting the case where a correct
integration is limited by, e.g. computational power.   
However, this estimator can be relative noisy because of its dependence on the value at the PDF maximum. One way to reduce this scatter could be to fit a gaussian to the PDF 
near its peak.  

\section{X-RAY GAS SHAPE AND CLUSTER MASS BIAS\label{massbias_section}}

By knowing the 3D x-ray gas temperature and density profiles one can
calculate the underlying total cluster density, and hence mass, by
combining the hydrostatic equilibrium (HE) equation 

\begin{equation}
\nabla(\rho_{gas}T_{gas})=-\rho_{gas}\nabla\Phi_{total}\label{eq:70}\end{equation}
with the poisson equation 

\begin{equation}
\nabla^{2}\Phi=4\pi G\rho_{total}\label{eq:80}\end{equation}
where the index \emph{'total'} indicates that the contribution is
from both gas and dark matter. 
When x-ray observations are possible for a cluster and the x-ray gas is in HE, this method is
one of the most precise ways to estimate the cluster mass as a function of radius within the visible x-ray region.
However, as we can see, the estimated cluster mass
will be wrong if the gas is not in HE or if $\rho_{gas}$, $T_{gas}$
is not correctly known. One way of misestimating $\rho_{gas}$ and $T_{gas}$
is fitting a spherical model to data for an intrinsic non-spherical
gas structure. Depending on the shape, this assumption will propagate
to a bias in the estimated total cluster mass. In this section we
will study the cluster mass bias as a function of different shapes
along the line of sight. 

\subsection{Mass Bias}

\begin{figure}[h!]
\centering
\includegraphics[scale=1.1]{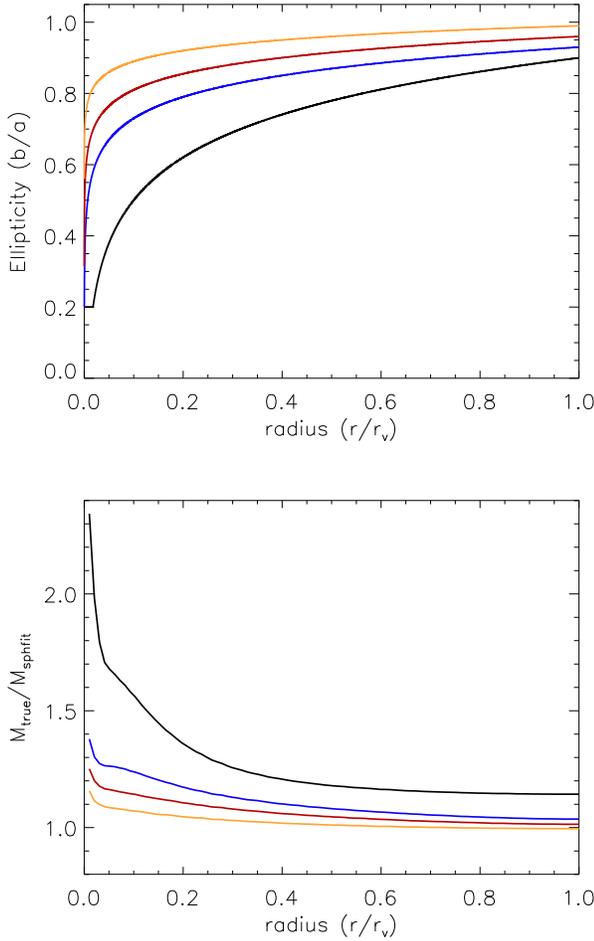}

\caption{\small{Mass bias from assuming spherical symmetry when fitting to
non-spherical x-ray structures. Upper: Shape along the line
of sight for four different prolate x-ray structures. Lower: Ratio between
the true total mass profile and the mass profile estimated from a
spherical fit to dataset 'shM2'  having the shapes shown at
the upper plot.
Opposite bias (i.e. $M_{true}/M_{sphfit}<1$) is expected for oblate structures.  
\label{fig:mass_bias}}}
 
\end{figure}

The upper plot in figure \ref{fig:mass_bias} shows the shape along
the line of sight for four different x-ray structures. We take the
four structures to have temperature and density profiles similar
to shM2, but different spacial shapes. Fitting temperature and density profiles to these four
structures assuming spherical symmetry, will result in biased mass
profiles. The ratio between the biased and the true mass profile is shown in
the lower plot in figure \ref{fig:mass_bias}. 
We have only included the mass contribution within $r_v$. 
Taking the rest of the mass of the
cluster into account, requires an extrapolation of the dark matter potential form
beyond the visible x-ray region. This is necessary when combining or comparing
with other mass probes such as lensing.

As seen on the plot, the shapes we are considering leads to small
biases at the $10$ percent level, dependent on the radius. This difference 
can be important for doing future precision cosmology
using clusters. However, at this level the degree of hydrostatic equilibrium
may lead to higher uncertainties in the mass estimation (\cite{2009ApJ...705.1129L}, \cite{2008A&A...491...71P}, \cite{2011A&A...525A.110C}).

\section{CONCLUSIONS}

We have presented a new method for measuring a radial dependent shape along the line of sight of the intracluster x-ray emitting gas. The method uses the assumption that
the shape, temperature and density profiles can be described by parameterized functions. Compared to several previous studies, we use the whole spectral information. 
Using this method we have demonstrated the possibilities for measuring shape on \verb+CHANDRA+ mock data. 

We find that around $10^{6}$ photons are required to get a $5-\sigma$ detection of shape when fitting to a model showing realistic features of the gas, such as cool core and a double powerlaw for the density profile. We have seen, by presenting correlations matrices, that density and shape have a strong correlation, whereas
temperature is essentially uncorrelated. This strong correlation indicates that independent measurements of the density profile can strongly improve the estimation of shape. 

We demonstrated that Bayes'  theorem very effectively can be used to compare different prior input models for our approach. This is of great importance since
the actual science one extracts in the end has to be read off from the input model. 

Finally we showed the effect on the mass profile estimation from assuming spherical symmetry when fitting structures with non-spherical shapes. 
Within our considered class of shapes, we found the mass estimation to be biased at the $10\%$ level. 

In a future paper we will use our framework on real data.

\acknowledgments

We warmly thank Martina Zamboni for useful discussions.
The Dark Cosmology Centre is funded by the Danish National Research Foundation.

\bibliographystyle{apj} 
\bibliography{xray_shape_submit}

\begin{thebibliography}{46}
\expandafter\ifx\csname natexlab\endcsname\relax\def\natexlab#1{#1}\fi

\bibitem[{{Allen} {et~al.}(2011){Allen}, {Evrard}, \&
  {Mantz}}]{2011ARA&A..49..409A}
{Allen}, S.~W., {Evrard}, A.~E., \& {Mantz}, A.~B. 2011, \araa, 49, 409

\bibitem[{{Allen} {et~al.}(2008){Allen}, {Rapetti}, {Schmidt}, {Ebeling},
  {Morris}, \& {Fabian}}]{2008MNRAS.383..879A}
{Allen}, S.~W., {Rapetti}, D.~A., {Schmidt}, R.~W., {Ebeling}, H., {Morris},
  R.~G., \& {Fabian}, A.~C. 2008, \mnras, 383, 879

\bibitem[{{Amanullah} {et~al.}(2011){Amanullah}, {Goobar}, {Cl{\'e}ment},
  {Cuby}, {Dahle}, {Dahl{\'e}n}, {Hjorth}, {Fabbro}, {J{\"o}nsson}, {Kneib},
  {Lidman}, {Limousin}, {M{\"o}rtsell}, {Nordin}, {Paech}, {Richard}, {Riehm},
  {Stanishev}, \& {Watson}}]{2011arXiv1109.4740A}
{Amanullah}, R., {et~al.} 2011, ArXiv e-prints

\bibitem[{{Andrae} {et~al.}(2010){Andrae}, {Schulze-Hartung}, \&
  {Melchior}}]{2010arXiv1012.3754A}
{Andrae}, R., {Schulze-Hartung}, T., \& {Melchior}, P. 2010, ArXiv e-prints

\bibitem[{{Arnaud}(1996)}]{1996ASPC..101...17A}
{Arnaud}, K.~A. 1996, in Astronomical Society of the Pacific Conference Series,
  Vol. 101, Astronomical Data Analysis Software and Systems V, ed.
  {G.~H.~Jacoby \& J.~Barnes}, 17--+

\bibitem[{{Bradley} {et~al.}(2011){Bradley}, {Bouwens}, {Zitrin}, {Smit},
  {Coe}, {Ford}, {Zheng}, {Illingworth}, {Ben{\'{\i}}tez}, \&
  {Broadhurst}}]{2011arXiv1104.2035B}
{Bradley}, L.~D., {et~al.} 2011, ArXiv e-prints

\bibitem[{{Cavaliere} {et~al.}(2011){Cavaliere}, {Lapi}, \&
  {Fusco-Femiano}}]{2011A&A...525A.110C}
{Cavaliere}, A., {Lapi}, A., \& {Fusco-Femiano}, R. 2011, \aap, 525, A110+

\bibitem[{Chib \& Greenberg(1995)}]{Chib:1995:UMH}
Chib, S., \& Greenberg, E. 1995, 49, 327

\bibitem[{{Chongchitnan} \& {Silk}(2011)}]{2011arXiv1107.5617C}
{Chongchitnan}, S., \& {Silk}, J. 2011, ArXiv e-prints

\bibitem[{{Conte} {et~al.}(2011){Conte}, {de Petris}, {Comis}, {Lamagna}, \&
  {de Gregori}}]{2011A&A...532A..14C}
{Conte}, A., {de Petris}, M., {Comis}, B., {Lamagna}, L., \& {de Gregori}, S.
  2011, \aap, 532, A14+

\bibitem[{{De Filippis} {et~al.}(2005){De Filippis}, {Sereno}, {Bautz}, \&
  {Longo}}]{2005ApJ...625..108D}
{De Filippis}, E., {Sereno}, M., {Bautz}, M.~W., \& {Longo}, G. 2005, \apj,
  625, 108

\bibitem[{{Fedeli} {et~al.}(2009){Fedeli}, {Moscardini}, \&
  {Matarrese}}]{2009MNRAS.397.1125F}
{Fedeli}, C., {Moscardini}, L., \& {Matarrese}, S. 2009, \mnras, 397, 1125

\bibitem[{{Fusco-Femiano} {et~al.}(2005){Fusco-Femiano}, {Landi}, \&
  {Orlandini}}]{2005ApJ...624L..69F}
{Fusco-Femiano}, R., {Landi}, R., \& {Orlandini}, M. 2005, \apjl, 624, L69

\bibitem[{{Hansen} \& {Piffaretti}(2007)}]{2007A&A...476L..37H}
{Hansen}, S.~H., \& {Piffaretti}, R. 2007, \aap, 476, L37

\bibitem[{{Hayashi} {et~al.}(2007){Hayashi}, {Navarro}, \&
  {Springel}}]{2007MNRAS.377...50H}
{Hayashi}, E., {Navarro}, J.~F., \& {Springel}, V. 2007, \mnras, 377, 50

\bibitem[{{Host} \& {Hansen}(2011)}]{2011ApJ...736...52H}
{Host}, O., \& {Hansen}, S.~H. 2011, \apj, 736, 52

\bibitem[{{Jain} \& {Zhang}(2008)}]{2008PhRvD..78f3503J}
{Jain}, B., \& {Zhang}, P. 2008, \prd, 78, 063503

\bibitem[{{Jenkins} \& {Peacock}(2011)}]{2011MNRAS.413.2895J}
{Jenkins}, C.~R., \& {Peacock}, J.~A. 2011, \mnras, 413, 2895

\bibitem[{{Kaastra} {et~al.}(2004){Kaastra}, {Tamura}, {Peterson}, {Bleeker},
  {Ferrigno}, {Kahn}, {Paerels}, {Piffaretti}, {Branduardi-Raymont}, \&
  {B{\"o}hringer}}]{2004A&A...413..415K}
{Kaastra}, J.~S., {et~al.} 2004, \aap, 413, 415

\bibitem[{{Kneib} {et~al.}(2004){Kneib}, {Ellis}, {Santos}, \&
  {Richard}}]{2004ApJ...607..697K}
{Kneib}, J.-P., {Ellis}, R.~S., {Santos}, M.~R., \& {Richard}, J. 2004, \apj,
  607, 697

\bibitem[{{Lau} {et~al.}(2009){Lau}, {Kravtsov}, \&
  {Nagai}}]{2009ApJ...705.1129L}
{Lau}, E.~T., {Kravtsov}, A.~V., \& {Nagai}, D. 2009, \apj, 705, 1129

\bibitem[{{Lau} {et~al.}(2011){Lau}, {Nagai}, {Kravtsov}, \&
  {Zentner}}]{2011ApJ...734...93L}
{Lau}, E.~T., {Nagai}, D., {Kravtsov}, A.~V., \& {Zentner}, A.~R. 2011, \apj,
  734, 93

\bibitem[{{Lemze} {et~al.}(2009){Lemze}, {Broadhurst}, {Rephaeli}, {Barkana},
  \& {Umetsu}}]{2009ApJ...701.1336L}
{Lemze}, D., {Broadhurst}, T., {Rephaeli}, Y., {Barkana}, R., \& {Umetsu}, K.
  2009, \apj, 701, 1336

\bibitem[{{{\L}okas} \& {Mamon}(2003)}]{2003MNRAS.343..401L}
{{\L}okas}, E.~L., \& {Mamon}, G.~A. 2003, \mnras, 343, 401

\bibitem[{{Mantz} {et~al.}(2010){Mantz}, {Allen}, {Rapetti}, \&
  {Ebeling}}]{2010MNRAS.406.1759M}
{Mantz}, A., {Allen}, S.~W., {Rapetti}, D., \& {Ebeling}, H. 2010, \mnras, 406,
  1759

\bibitem[{{Morandi} {et~al.}(2011){Morandi}, {Limousin}, {Rephaeli}, {Umetsu},
  {Barkana}, {Broadhurst}, \& {Dahle}}]{2011MNRAS.416.2567M}
{Morandi}, A., {Limousin}, M., {Rephaeli}, Y., {Umetsu}, K., {Barkana}, R.,
  {Broadhurst}, T., \& {Dahle}, H. 2011, \mnras, 416, 2567

\bibitem[{{Morandi} {et~al.}(2010){Morandi}, {Pedersen}, \&
  {Limousin}}]{2010ApJ...713..491M}
{Morandi}, A., {Pedersen}, K., \& {Limousin}, M. 2010, \apj, 713, 491

\bibitem[{{Piffaretti} {et~al.}(2005){Piffaretti}, {Jetzer}, {Kaastra}, \&
  {Tamura}}]{2005A&A...433..101P}
{Piffaretti}, R., {Jetzer}, P., {Kaastra}, J.~S., \& {Tamura}, T. 2005, \aap,
  433, 101

\bibitem[{{Piffaretti} \& {Valdarnini}(2008)}]{2008A&A...491...71P}
{Piffaretti}, R., \& {Valdarnini}, R. 2008, \aap, 491, 71

\bibitem[{{Planck Collaboration} {et~al.}(2011){Planck Collaboration},
  {Aghanim}, {Arnaud}, {Ashdown}, {Aumont}, {Baccigalupi}, {Balbi}, {Banday},
  {Barreiro}, {Bartelmann}, \& et~al.}]{2011arXiv1101.2043P}
{Planck Collaboration} {et~al.} 2011, ArXiv e-prints

\bibitem[{{Pointecouteau} {et~al.}(2005){Pointecouteau}, {Arnaud}, \&
  {Pratt}}]{2005A&A...435....1P}
{Pointecouteau}, E., {Arnaud}, M., \& {Pratt}, G.~W. 2005, \aap, 435, 1

\bibitem[{{Postman} {et~al.}(2011){Postman}, {Coe}, {Benitez}, {Bradley},
  {Broadhurst}, {Donahue}, {Ford}, {Graur}, {Graves}, {Jouvel}, {Koekemoer},
  {Lemze}, {Medezinski}, {Molino}, {Moustakas}, {Ogaz}, {Riess}, {Rodney},
  {Rosati}, {Umetsu}, {Zheng}, {Zitrin}, {Bartelmann}, {Bouwens}, {Host},
  {Infante}, {Jha}, {Jimenez-Teja}, {Kelson}, {Lahav}, {Lazkoz}, {Maoz},
  {McCully}, {Melchior}, {Meneghetti}, {Merten}, {Nonino}, {Patel}, {Regos},
  {Seitz}, {Sayers}, {Golwala}, \& {Van der Wel}}]{2011arXiv1106.3328P}
{Postman}, M., {et~al.} 2011, ArXiv e-prints

\bibitem[{{Rapetti} {et~al.}(2010){Rapetti}, {Allen}, {Mantz}, \&
  {Ebeling}}]{2010MNRAS.406.1796R}
{Rapetti}, D., {Allen}, S.~W., {Mantz}, A., \& {Ebeling}, H. 2010, \mnras, 406,
  1796

\bibitem[{{Sarazin}(1988)}]{1988xrec.book.....S}
{Sarazin}, C.~L. 1988, {X-ray emission from clusters of galaxies}, ed.
  {Sarazin, C.~L.}

\bibitem[{{Sartoris} {et~al.}(2010){Sartoris}, {Borgani}, {Fedeli},
  {Matarrese}, {Moscardini}, {Rosati}, \& {Weller}}]{2010MNRAS.407.2339S}
{Sartoris}, B., {Borgani}, S., {Fedeli}, C., {Matarrese}, S., {Moscardini}, L.,
  {Rosati}, P., \& {Weller}, J. 2010, \mnras, 407, 2339

\bibitem[{{Schafer}(1991)}]{1991xxrs.book.....S}
{Schafer}, R.~A. 1991, {XSPEC, an x-ray spectral fitting package : version 2 of
  the user's guide}, ed. {Schafer, R.~A.}

\bibitem[{{Sereno} {et~al.}(2011){Sereno}, {Ettori}, \&
  {Baldi}}]{2011arXiv1109.2732S}
{Sereno}, M., {Ettori}, S., \& {Baldi}, A. 2011, ArXiv e-prints

\bibitem[{{Sereno} \& {Umetsu}(2011)}]{2011MNRAS.416.3187S}
{Sereno}, M., \& {Umetsu}, K. 2011, \mnras, 416, 3187

\bibitem[{{Stark} {et~al.}(2007){Stark}, {Ellis}, {Richard}, {Kneib}, {Smith},
  \& {Santos}}]{2007ApJ...663...10S}
{Stark}, D.~P., {Ellis}, R.~S., {Richard}, J., {Kneib}, J.-P., {Smith}, G.~P.,
  \& {Santos}, M.~R. 2007, \apj, 663, 10

\bibitem[{{Trotta}(2008)}]{2008ConPh..49...71T}
{Trotta}, R. 2008, Contemporary Physics, 49, 71

\bibitem[{{Vikhlinin} {et~al.}(2006){Vikhlinin}, {Kravtsov}, {Forman}, {Jones},
  {Markevitch}, {Murray}, \& {Van Speybroeck}}]{2006ApJ...640..691V}
{Vikhlinin}, A., {Kravtsov}, A., {Forman}, W., {Jones}, C., {Markevitch}, M.,
  {Murray}, S.~S., \& {Van Speybroeck}, L. 2006, \apj, 640, 691

\bibitem[{{Vikhlinin} {et~al.}(2009{\natexlab{a}}){Vikhlinin}, {Kravtsov},
  {Burenin}, {Ebeling}, {Forman}, {Hornstrup}, {Jones}, {Murray}, {Nagai},
  {Quintana}, \& {Voevodkin}}]{2009ApJ...692.1060V}
{Vikhlinin}, A., {et~al.} 2009{\natexlab{a}}, \apj, 692, 1060

\bibitem[{{Vikhlinin} {et~al.}(2009{\natexlab{b}}){Vikhlinin}, {Murray},
  {Gilli}, {Tozzi}, {Paolillo}, {Brandt}, {Tagliaferri}, {Bautz}, {Allen},
  {Donahue}, {Evrad}, {Flanagan}, {Rosati}, {Borgani}, {Giacconi}, {Weisskopf},
  {Ptak}, {Alexander}, {Pareschi}, {Forman}, \& {Jones}}]{2009astro2010S.305V}
{Vikhlinin}, A., {et~al.} 2009{\natexlab{b}}, in ArXiv Astrophysics e-prints,
  Vol. 2010, astro2010: The Astronomy and Astrophysics Decadal Survey, 305--+

\bibitem[{{Wojtak} {et~al.}(2011){Wojtak}, {Hansen}, \&
  {Hjorth}}]{2011arXiv1109.6571W}
{Wojtak}, R., {Hansen}, S.~H., \& {Hjorth}, J. 2011, ArXiv e-prints

\bibitem[{{Wojtak} \& {{\L}okas}(2010)}]{2010MNRAS.408.2442W}
{Wojtak}, R., \& {{\L}okas}, E.~L. 2010, \mnras, 408, 2442

\bibitem[{{Yoo} {et~al.}(2009){Yoo}, {Fitzpatrick}, \&
  {Zaldarriaga}}]{2009PhRvD..80h3514Y}
{Yoo}, J., {Fitzpatrick}, A.~L., \& {Zaldarriaga}, M. 2009, \prd, 80, 083514

\end{thebibliography}

\appendix

\section{APPENDIX}

\subsubsection{Creating artificial observations of an x-ray gas\label{sub:Creating-artificial-observations}}

In our analysis we have two different situations where we need to
simulate a dataset. The first is as input to the MCMC routine when
fitting to a given dataset. The second is where we actually simulate
the dataset that has to be fitted, i.e. the mock data. The first steps for both are the
same, and is described in the following; Given a set of parameterized
profiles and shape we create a three dimensional x-ray gas on a grid.
The local spectral information is calculated by \verb+XSPEC+'s (see e.g. \cite{1996ASPC..101...17A}, \cite{1991xxrs.book.....S}) 
model mekal (http://heasarc.nasa.gov/xanadu/\verb+XSPEC+/manual/XSmodelMekal.html and references within) at redshift zero including galactic absorption. We use five
times higher spatial resolution in the inner regions compared to the
outer parts, to make sure no resolution effects propagate into the
results. We then project all the spectral information onto the 2D
observational plane defined such that the x-gas structure is spherical
symmetric in that plane. The projected data is then convolved
in \verb+XSPEC+ with an instrumental response function, here chosen to be
from \verb+CHANDRA+, to create a final observed picture. In an ideal world
this is the picture read out from the instrument assuming pixelation
from the CCD is unimportant. In a real world, a spacial
and spectral rebinning is done at this step. When we create a dataset as input to the MCMC routine, 
the binning is done so that it matches the binning
of the observed dataset. When generating a mock dataset we do the binning such that 
the radial bins have the same number photon counts and the spectral
bins have more than a given threshold. This ensures equally statistical
weights for each bin. For the fits in this paper, we fixed the number
of radial bins to 12 for all datasets. Because an x-ray gas density
profile usually have a logarithmic shape, the radial bins are therefore
also approximately logarithmic linear spaced. Our spectral threshold
is chosen such that the number of new spectral bins are around 200,
of originally 1024. This corresponds to a threshold of 20 counts
per spectral bin for a $6\cdot10^{4}$ number photons observation.
It was not computationally possible to scan over different binning
strategies, but the chosen binning is believed to match a real
case scenario fairly well. 

\begin{figure}[h!]
\includegraphics[scale=0.30]{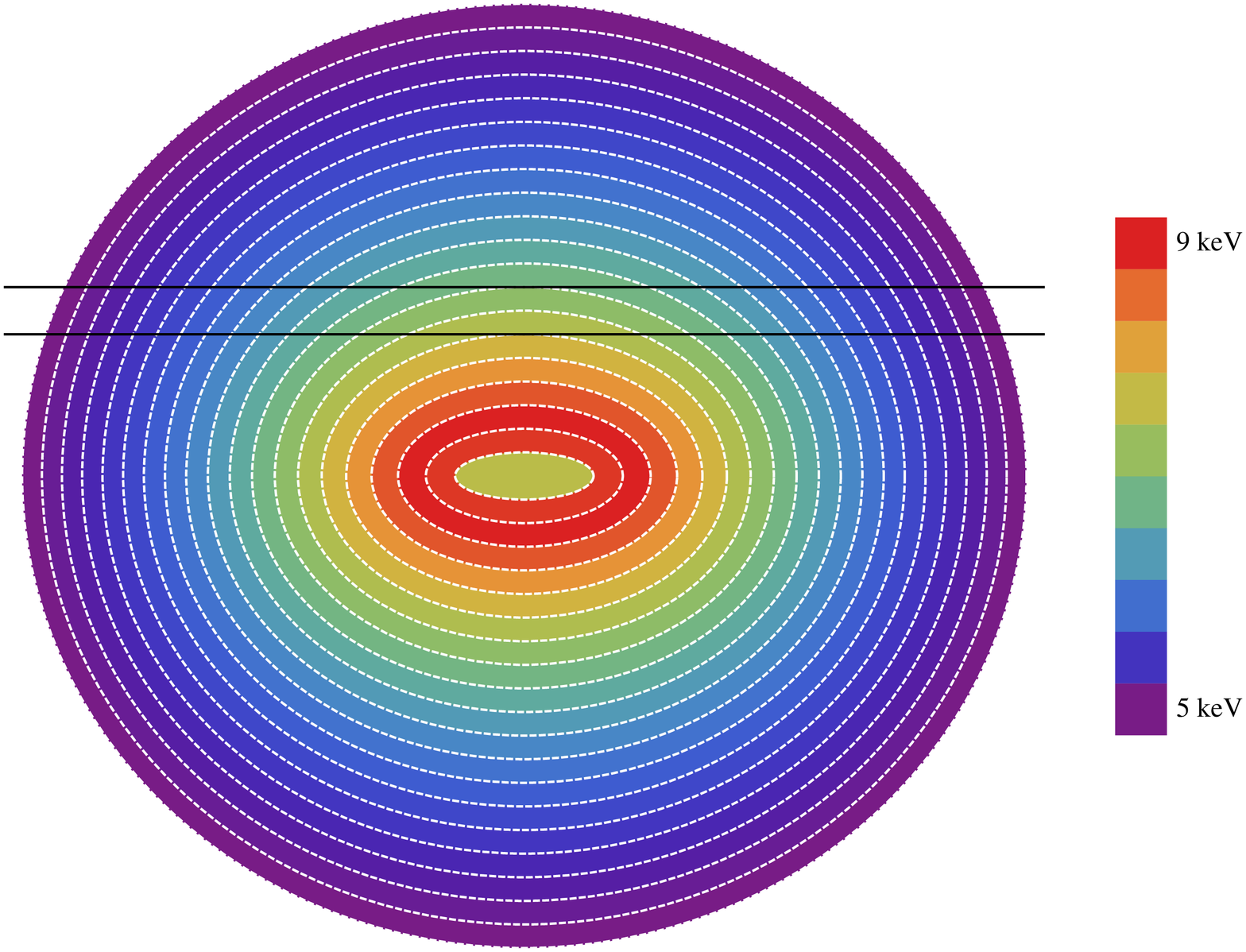}\includegraphics[scale=0.42]{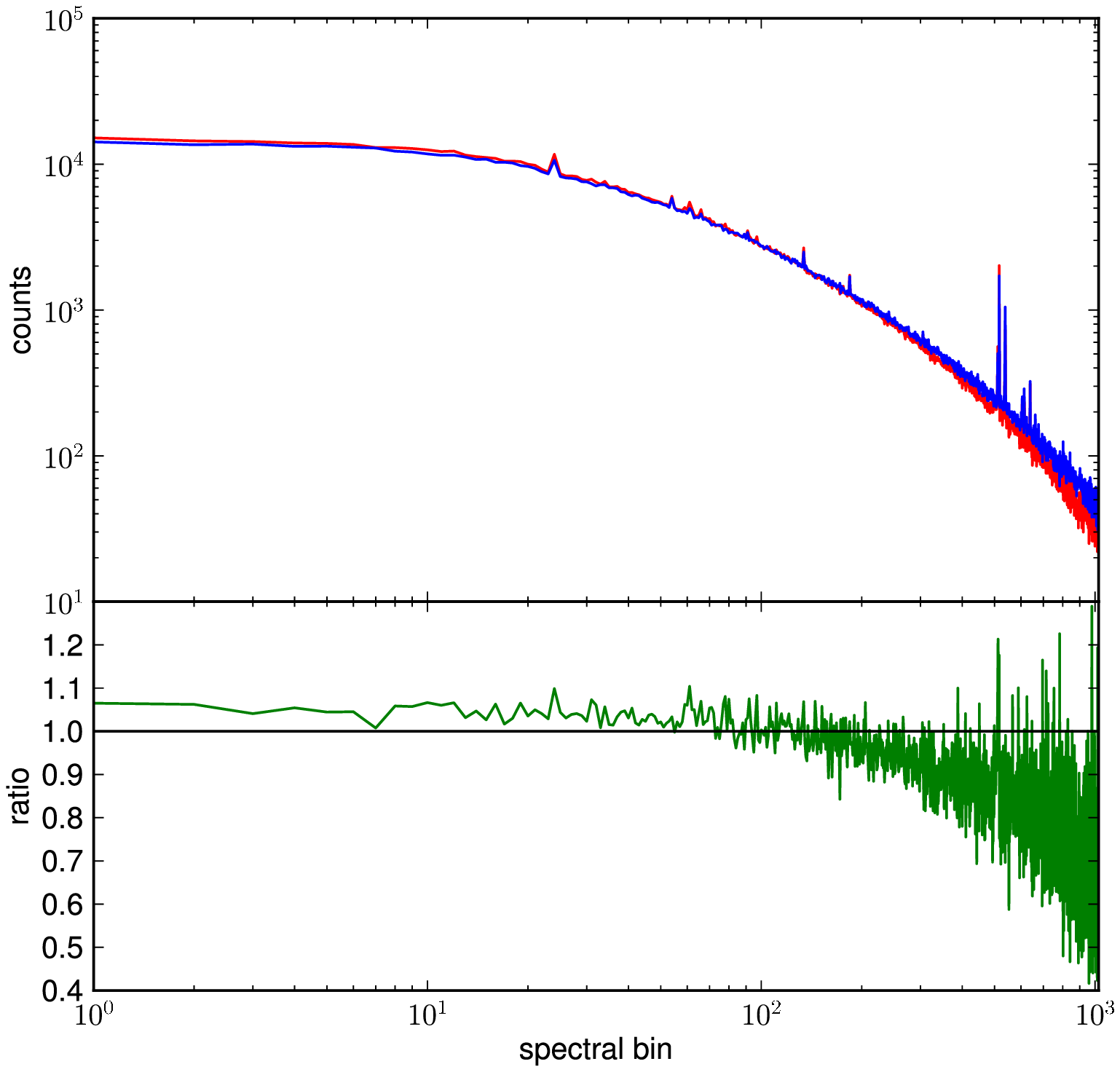}

\caption{\small{Left: A generated x-ray gas map with no noise added in the plane along the line of sight for structure 'shM2' introduced in section \ref{sub:A-more-realistic}. The color indicates the temperature in keV.
Right: Upper plot shows two spectra generated from the region between the two black lines shown in the left plot. 
The spectrum in red is a free-free spectrum generated with the mean x-ray temperature from the region between the two black lines in the left plot and the spectrum in blue is the true projected spectrum,
 i.e. the sum of many free-free spectra each generated locally in the x-gas. The lower plot shows the ratio between the red and the blue spectra.
\label{fig:gas_and_spectra}}}

\end{figure}

Figure \ref{fig:gas_and_spectra} (left) illustrates a noise free generated x-ray gas map with a non-spherical shape and its temperature profile. The shape and the
temperature profile is the one used for 'shM2' introduced in section \ref{sub:A-more-realistic}. The right plot in figure \ref{fig:gas_and_spectra} shows two spectra generated from the region between the two black lines
shown in the left plot. The spectrum in red is a free-free spectrum generated with \verb+XSPEC+ using the mean projected temperature 
and the spectrum in blue is the true projected spectrum, i.e. the sum of many free-free spectra each generated locally in the x-gas. The difference seen in the lower part of the right plot is basically what give us information about shape
and profiles.

\subsubsection{Monte Carlo Technics used for this paper\label{sub:Monte-Carlo-Technics}}

We wrote a Monte Carlo Markov-Chain (MCMC) algorithm for fitting to
a data set. The MCMC uses a Metropolis-Hastings sampling (\cite{Chib:1995:UMH})
with a flat and symmetric proposal density. The size of this proposal
density was tuned to reach an acceptance rate of around 0.2-0.3 which
has been shown to be the most optimal for sampling higher dimensional
distributions. The width of the proposal density along each
parameter axes was tuned in units of the root mean square for the
individual PDF for each parameter. For all runs the sampling space
was limited by bounds on each parameter axes and realizations with
a temperature profile exceeding 15 kev or going below 0.5 kev was
given zero probability. Among numerous tests of possible resolution, boundary or sampling
effects we tested that the codes reproduced the theoretical expected
degeneracy between an overall density scaling and a fixed axis ratio
along the line of sight. We tested this up to a total number of 500.000 photons.
A sample of tests was also done against
an independently written code which generates artificial x-ray data
using the {}``shell binning'' approach (see e.g. http://cxc.harvard.edu/contrib/deproject/). We tested convergence
by starting chains at random places and with different scalings (number
photons) of the PDF. All distributions shown in the paper are based
on $5\cdot10^{6}$ samplings. The fitting results presented are based
on one realization of data, marginalizing over several realizations
was not computationally possible. We assumed a diagonal covariance
matrix for the observed photon measurements and the noise to be gaussian.

\end{document}